\documentclass[
 ,final            
  ]
  {aipproc}

\layoutstyle{6x9}

\begin{document}

\title{Compton Scattering and Generalized
Polarizabilities}

\author{Stefan Scherer}{
  address={Institut f\"ur Kernphysik, Johannes Gutenberg-Universit\"at, 
  D-55099 Mainz, Germany}
}

\begin{abstract}
   In recent years, real and virtual Compton scattering off the nucleon
have attracted considerable interest from both the experimental and theoretical
sides.
   Real Compton scattering gives access to the so-called electromagnetic
polarizabilities containing the structure information beyond the global
properties of the nucleon such as its charge, mass, and magnetic moment.
   These polarizabilities have an intuitive interpretation in terms
of induced dipole moments and thus characterize the response of the constituents
of the nucleon to a soft external stimulus.
   The virtual Compton scattering reaction $e^-p\to e^-p\gamma$ allows
one to map out the {\em local} response to external fields and can be described
in terms of generalized electromagnetic polarizabilities.
   A simple classical interpretation in terms of the induced electric
and magnetic polarization densities is proposed.
   We will discuss experimental results for the polarizabilities of the
proton and compare them with theoretical predictions.
\end{abstract}

\maketitle


\section{Introduction and Overview}

    Real Compton scattering (RCS), $\gamma(q,\epsilon(\lambda))+N(p,s) \to
\gamma(q',\epsilon'(\lambda'))+N(p',s')$, has a long history of providing
important theoretical and experimental tests for models of nucleon structure
(see, {\em e.g.}, Refs. \cite{Lvov:1993fp,Scherer:1999yw,Drechsel:2002ar} for an
introduction).
   Based on the requirement of gauge invariance, Lorentz invariance,
crossing symmetry, and the discrete symmetries, the famous low-energy theorem of
Low \cite{Low:1954kd} and Gell-Mann and Goldberger \cite{Gell-Mann:1954kc}
uniquely specifies the terms in the low-energy scattering amplitude up to and
including terms linear in the photon momentum.
   The coefficients of this expansion are expressed in terms of global
properties of the nucleon: its mass, charge, and magnetic moment.
   In principle, any model respecting the symmetries entering the derivation
of the LET should reproduce the constraints of the LET. It is only terms of
second order which contain new information on the structure of the nucleon
specific to Compton scattering.
   For a general target, these effects can be parameterized in terms of
two constants, the electric and magnetic polarizabilities $\alpha$ and $\beta$,
respectively \cite{Klein:1955}.

   The scattering amplitude may be parameterized in terms of six independent
functions $A_i$ depending on the photon energy $\omega$ and the scattering angle,
\begin{equation}
\label{trcs} T=\vec{\epsilon}'^{\ast}\cdot\vec{\epsilon}\, A_1
+\vec{\epsilon}'^\ast\cdot\hat{q}\,\vec{\epsilon}\cdot\hat{q}'\, A_2
+i\vec{\sigma}\cdot \vec{\epsilon}'^{\ast}\times\vec{\epsilon}\, A_3 +\cdots.
\end{equation}
   In the forward and backward directions only two functions, namely,
$A_1$ and $A_3$, contribute.
   For example, the Taylor series expansion of $A_1$, for the proton, is given
by
\begin{equation}
A_1 = -\frac{e^2}{m} + 4\pi(\alpha+\beta z)\omega^2 - \frac{e
^2}{4m^3}(1-z)\omega^2 + [\omega^4]\ , \label{A1}
\end{equation}
where $z=\cos(\theta)$.
   The leading-order term is given by the Thomson term and
the forward and backward amplitudes are sensitive to the combinations
$\alpha+\beta$ and $\alpha-\beta$, respectively.
   The sum of the polarizabilities is constrained by the Baldin sum rule
\cite{Baldin:1960},
\begin{equation}
\label{baldin} \alpha+\beta=\frac{1}{2\pi^2}\int_{\omega_{\rm thr}}^\infty
\frac{\sigma^{\rm tot}(\omega)}{\omega^2}d\omega,
\end{equation}
where $\sigma^{\rm tot}(\omega)$ is the total photoabsorption cross section.
   A fit to all modern low-energy experiments
\cite{Federspiel:1991yd,Zieger:1992jq,MacGibbon:1995in,OlmosdeLeon:2001zn} and
the sum rule relation $\alpha_p+\beta_p = (13.8\pm 0.4)$ \footnote{The
polarizabilities $\alpha$ and $\beta$ are given in units of
$10^{-4}\,\mbox{fm}^3$.} leads to the result \cite{OlmosdeLeon:2001zn}
\begin{displaymath}
\alpha_p=12.1 \pm 0.3_{\rm stat.} \mp 0.4_{\rm syst.} \pm 0.3_{\rm mod.},\quad
\beta_p=1.6 \pm 0.4_{\rm stat.} \pm 0.4_{\rm syst.} \pm 0.4_{\rm mod.}.
\end{displaymath}
   Clearly, the electric polarizability $\alpha_p$ dominates over the small magnetic
polarizability $\beta_p$, the smallness of which is thought to result from a
delicate cancellation between the paramagnetic $\Delta$ contribution and a nearly
equally strong diamagnetic term.
   Although, for example, soliton models of the nucleon have predicted such
a destructive interference for a long time \cite{Scoccola:1990yh,Scherer:1992jb},
the precise microscopic origin of the relatively large diamagnetic contribution
is still under debate.

   Information on the neutron has been obtained via low-energy neutron-$^{208}$Pb scattering
\cite{Schmiedmayer:1991},
\begin{displaymath}
\alpha_n=12.0\pm 1.5_{\rm stat.} \pm 2.0_{\rm syst.},
\end{displaymath}
quasi-free Compton scattering $\gamma d\to \gamma'np$ \cite{Kossert:2002jc},
\begin{displaymath}
\alpha_n = 12.5\pm 1.8_{\rm stat.}\,{^{+1.1}_{-0.6}}_{\rm syst.} \pm 1.1_{\rm
mod.},\quad \beta_n =  2.7\mp 1.8_{\rm stat.}\,{^{+0.6}_{-1.1}}_{\rm syst.} \mp
1.1_{\rm  mod.},
\end{displaymath}
and elastic $\gamma d$ scattering \cite{Lundin:2002jy}
\begin{displaymath}
\alpha_n=8.8\pm 2.4_{\rm stat.+syst.} \pm 3.0_{\rm mod.},\quad \beta_n=6.5\mp
2.4_{\rm stat.+syst.} \mp 3.0_{\rm mod.}.
\end{displaymath}
  A recent re-analysis of the sum rule yields $\alpha_n+\beta_n=(15.2\pm 0.5)$
\cite{Levchuk:1999zy}.

   New extractions using effective field theory for the nucleon
as well as the deuteron have yielded for the proton polarizabilities at ${\cal
O}(p^4)$ in chiral perturbation theory \cite{Beane:2004ra}
\begin{displaymath}
\alpha_p=12.1\pm 1.1_{\rm stat.} \pm 0.5_{\rm mod.},\quad \beta_p=3.4\pm 1.1_{\rm
stat.} \pm 0.1_{\rm mod.},
\end{displaymath}
and for the isoscalar nucleon polarizabilities
\begin{displaymath}
\alpha_N=13.0\pm 1.9_{\rm stat.}\,{^{+3.9}_{-1.5}}_{\rm mod.},
\end{displaymath}
and a $\beta_N$ that is consistent with zero within sizeable error bars.
   Another calculation including explicit $\Delta$ degrees of freedom obtained
\cite{Hildebrandt:2003fm,Hildebrandt:2004hh}
\begin{eqnarray*}
\alpha_p&=&11.04\pm 1.36,\quad
\beta_p=2.76\mp 1.36,\\
\alpha_N&=&12.6\pm 1.4_{\rm stat.}\pm 1.9_{\rm syst.},\quad \beta_N=2.3\pm
1.7_{\rm stat.}\pm 0.3_{\rm syst.}.
\end{eqnarray*}
 For a recent review on the status of the spin polarizabilities, see
Ref.\ \cite{Drechsel:2002ar}.

\section{Virtual Compton Scattering and Generalized Polarizabilities}
   As in all studies with electromagnetic probes, the possibilities to
investigate the structure of the target are much greater if virtual photons are
used, since the energy and the three-momentum of the virtual photon can be varied
independently.
   Moreover, the longitudinal component of current operators entering
the amplitude can be studied.
   The amplitude for virtual Compton scattering (VCS) off the proton,
$T_{\rm VCS}$, is accessible in the reaction $e^-p\to e^-p\gamma$.
   Similarly to Eq.\ (\ref{trcs}), $T_{\rm VCS}$ can be expressed in terms of
eight transverse and four longitudinal amplitudes.
   Model-independent predictions, 
based on Lorentz invariance, gauge invariance, crossing symmetry, and the
discrete symmetries, have been derived in Ref.\ \cite{Scherer:1996ux}.
  Up to and including terms of second order in the momenta $|\vec{q}\,|$ and
$|\vec{q}\,'|$, all functions $A_i$ are completely specified in terms of
quantities which can be obtained from elastic electron-proton scattering and RCS,
namely  $m$, $\kappa$, $G_E$, $G_M$, $r^2_E$, $\alpha$, and $\beta$.
   After dividing the amplitude $T_{\rm VCS}$ into a gauge-invariant
generalized pole piece $T_{\rm pole}$ and a residual piece $T_{\rm R}$, the
so-called generalized polarizabilities (GPs) of Ref.\ \cite{Guichon:1995pu}
result from an analysis of the residual piece in terms of electromagnetic
multipoles. A restriction to the lowest-order, {\em i.e.} linear terms in
$\omega'$ leads to only electric and magnetic dipole radiation in the final
state.
   Parity and angular-momentum selection rules, charge-conjugation symmetry,
and particle crossing generate six independent GPs
\cite{Guichon:1995pu,Drechsel:1996ag,Drechsel:1997xv}.

\begin{figure}[htbp]
\begin{minipage}[b]{0.5\textwidth}
\includegraphics[width=\textwidth]{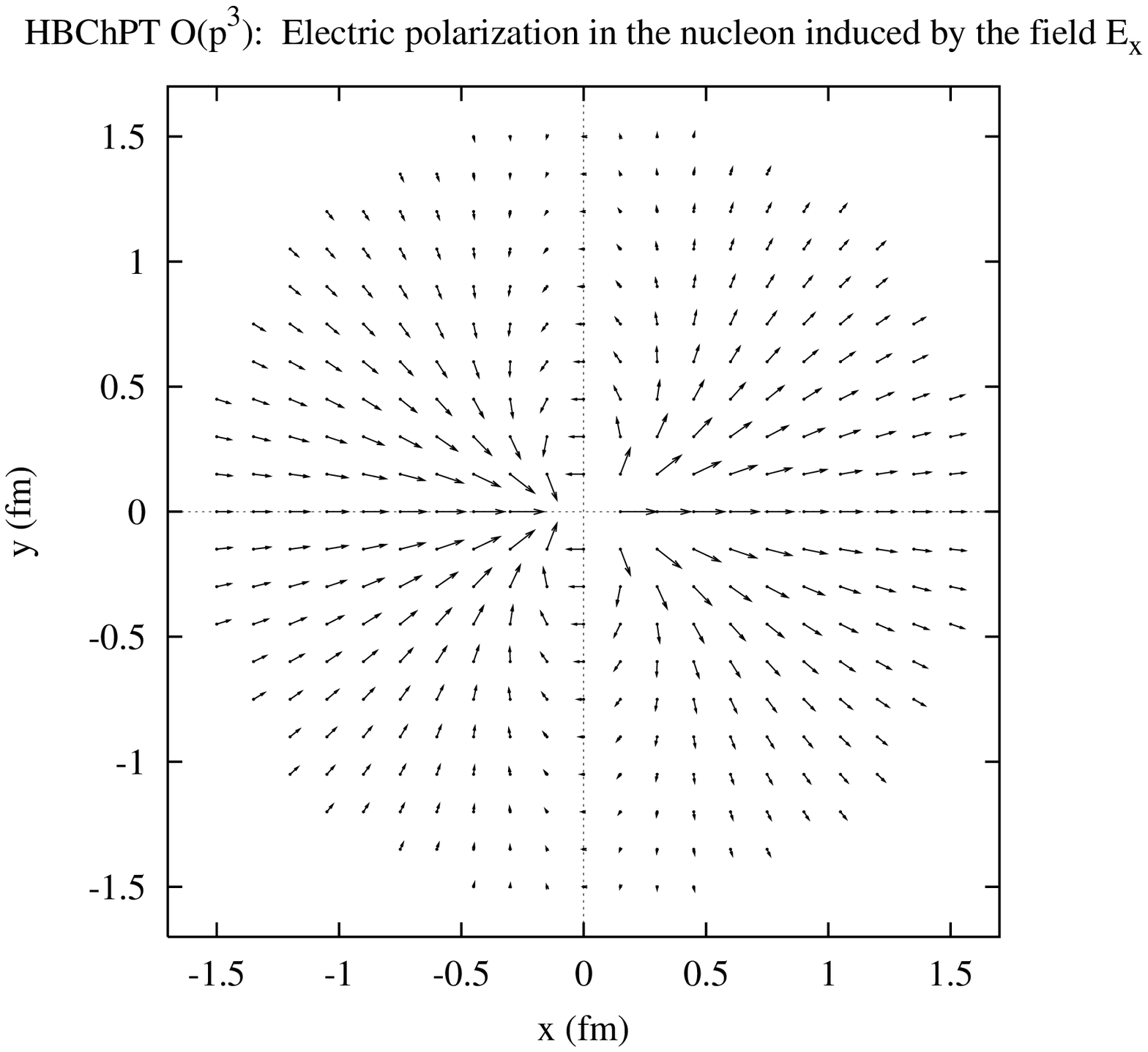}
\end{minipage}
\hspace{2em}
\begin{minipage}[b]{0.4\textwidth}
\includegraphics[width=\textwidth]{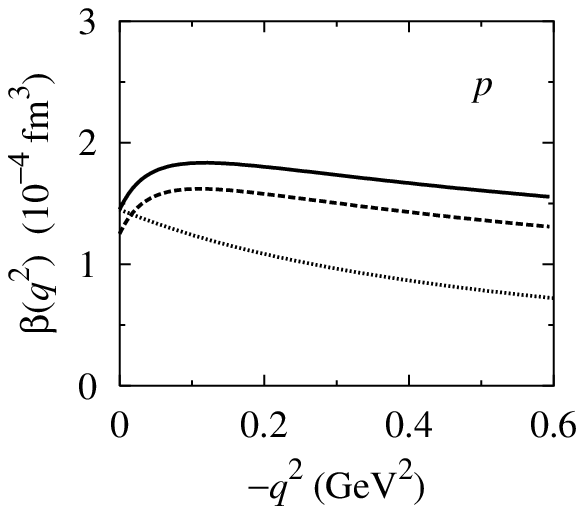}
\includegraphics[width=\textwidth]{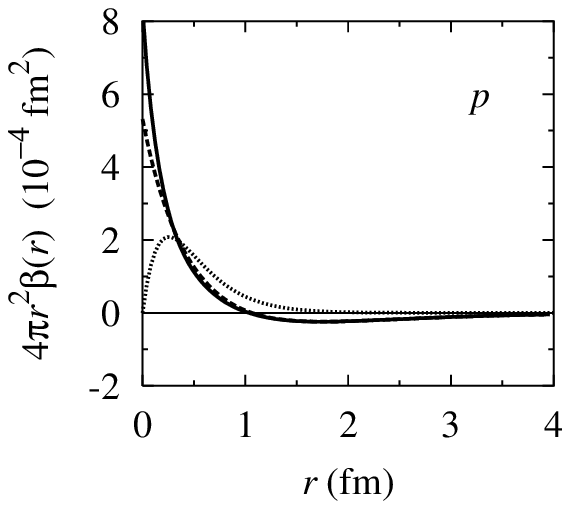}
\end{minipage}
\caption{\label{fig_gps} Left panel: Scaled electric polarization
$r^3 \alpha_{i1}$ [10$^{-3}$ fm$^3$] \cite{Lvov:2004}. Right panel: Generalized
magnetic polarizability $\beta(q^2)$ and density of magnetic polarizability
$\beta(r)$. Dashed lines: contribution of pion loops; solid lines: total
contribution; dotted lines: VMD predictions normalized to ${\beta}(0)$
\cite{L'vov:2001fz}.}
\end{figure}

    Similarly as elastic electron scattering allows one to map out the
spatial distribution of charge and magnetization inside the nucleon, the
generalized polarizabilities parameterize a {\em local} response of a system in
an external field.
   For example, if the nucleon is exposed to a static and uniform external
electric field  $\vec{E}$, an electric polarization $\vec{\cal P}$ is generated
which is related to the {\em density} of the induced electric dipole moments,
\begin{equation}
\label{H1:b:d-induced} {\cal P}_i(\vec r) = 4\pi\alpha_{ij}(\vec r)\,E_j.
\end{equation}
   The tensor $\alpha_{ij}(\vec r)$, {\em i.e.}~the density of the full electric
polarizability of the system, can be expressed as \cite{L'vov:2001fz}
\begin{displaymath}
\alpha_{ij}(\vec r) =
     \alpha_L(r) \hat r_i \hat r_j
      + \alpha_T(r) (\delta_{ij} - \hat r_i \hat r_j)
      + \frac{3\hat r_i \hat r_j - \delta_{ij}}{r^3}
        \int_r^\infty [\alpha_L(r')-\alpha_T(r')]\,r'^2\,dr',
\end{displaymath}
   where $\alpha_L(r)$ and $\alpha_T(r)$ are Fourier transforms
of the generalized longitudinal and transverse electric polarizabilities
$\alpha_L(q)$ and $\alpha_T(q)$, respectively.
   The definition of the generalized dipole polarizabilities of Ref.\ \cite{L'vov:2001fz}
has been obtained from a fully covariant framework as opposed to the multipole
decomposition of Ref.\ \cite{Guichon:1995pu}.
    In particular, it is important to realize that both longitudinal and
transverse polarizabilities are needed to fully recover the electric polarization
$\vec{\cal P}$.
   The left panel of Fig.\ \ref{fig_gps} shows the induced polarization inside a nucleon
as calculated in the framework of heavy-baryon chiral perturbation theory at
${\cal O}(p^3)$ \cite{Lvov:2004} and clearly shows that the polarization, in
general, does not point into the direction of the applied electric field.

   Similar considerations apply to an external magnetic field.
   Since the magnetic induction is always transverse
(i.e., $\vec\nabla\cdot\vec B=0$), it is sufficient to consider $\beta_{ij}(\vec
r)=\beta(r)\delta_{ij}$ \cite{L'vov:2001fz}.
   Then the magnetization $\vec{\cal
M}$ induced by the uniform external magnetic field is given in terms of the
density of the magnetic polarizability as $\vec{\cal M}(\vec r) =
4\pi\beta(r)\vec B$ (see right panel of Fig.\ \ref{fig_gps}).

   The first results for the two structure functions $P_{LL}-P_{TT}/\epsilon$
and $P_{LT}$ at $Q^2=0.33$ GeV$^2$ were obtained from a dedicated VCS experiment
at MAMI \cite{Roche:2000ng}.
   Results at higher four-momentum transfer squared
$Q^2=0.92$ and $Q^2=1.76$ GeV$^2$ have been reported in Ref.\
\cite{Laveissiere:2003sv}.
   Additional data are expected from MIT/Bates for $Q^2 =
0.05$ GeV$^2$ aiming at an extraction of the magnetic polarizability
\cite{Miskimen:2004}.
   Moreover, data in the resonance region have been taken at JLab
for $Q^2=1$ GeV$^2$ \cite{Fonvieille:2002fb} which have been analyzed in the
framework of the dispersion relation formalism of Ref.\
\cite{Drechsel:2002ar,Pasquini:2001yy}.

   Table \ref{H1:b2:tableresults} shows the experimental results of
\cite{Roche:2000ng} in combination with various model calculations.
   Clearly, the experimental precision of \cite{Roche:2000ng} already
allows for a critical test of the different models.
   Within ChPT and the linear sigma model, the GPs are essentially
due to pionic degrees of freedom.
   Due to the small pion mass the effect in the spatial distributions
extends to larger distances (see also right panel of Fig.\ \ref{fig_gps}).
   On the other hand, the constituent quark model and other phenomenological
models involving Gau{\ss} or dipole form factors typically show a faster decrease in
the range $Q^2 < 1$ GeV$^2$.

\begin{table}
\begin{tabular}{|c|c|c|}
\hline &$P_{LL}-P_{TT}/\epsilon$ $[\mbox{GeV}^{-2}]$ &
$P_{LT}$ $[\mbox{GeV}^{-2}]$\\
\hline Experiment \cite{Roche:2000ng} & $23.7\pm 2.2_{\rm stat.}\pm 4.3_{\rm
syst.} \pm 0.6_{\rm syst.norm.}$ &
 $-5.0
\pm 0.8_{\rm stat.} \pm 1.4_{\rm syst.} \pm 1.1_{\rm syst. norm.}$
\\
\hline
LSM \cite{Metz:1996fn} & 11.5 & 0.0\\
\hline
ELM \cite{Vanderhaeghen:1996iz} & 5.9 & $-1.9$\\
\hline
HBChPT \cite{Hemmert:1997at} & 26.0 & $-5.3$\\
\hline
NRCQM \cite{Pasquini:2000ue} & $19.2|14.9^\ast$ & $-3.2|-4.5^\ast$\\
\hline
\end{tabular}
\caption{\label{H1:b2:tableresults}
Experimental results and theoretical predictions for the structure functions
$P_{LL}-P_{TT}/\epsilon$ and $P_{LT}$ at $Q^2=0.33$ GeV$^2$ and $\epsilon=0.62$.
 $\ast$ makes use of symmetry under particle crossing and charge conjugation
which is not a symmetry of NRCQM. }
\end{table}

\bibliographystyle{aipproc}   

\hyphenation{Post-Script Sprin-ger}

\end{document}